\documentclass[journal=jacsat,manuscript=article]{achemso}

\usepackage[version=3]{mhchem} 
\usepackage{array}
\newcolumntype{P}[1]{>{\centering\arraybackslash}p{#1}}
\usepackage{tikz}
 
\author{Alessandro Pianelli}
\affiliation{Faculty of Engineering and Natural Sciences, Tampere University, Tampere 33720, Finland}
\author{Rakesh Dhama}
\affiliation{Faculty of Engineering and Natural Sciences, Tampere University, Tampere 33720, Finland}
\author{Jarosław Judek}
\affiliation{
Institute of Microelectronics and Optoelectronics, Warsaw University of Technology, Koszykowa 75, Warsaw 00-662, Poland}
\author{Rafał Mazur}
\affiliation{
Institute of Applied Physics, Military University of Technology, Warsaw 00-908, Poland}
\author{Humeyra Caglayan}
\email{humeyra.caglayan@tuni.fi}
\affiliation{Faculty of Engineering and Natural Sciences, Tampere University, Tampere 33720, Finland}

\title[An \textsf{achemso} demo]
  {Two-color all-optical switching in Si-compatible epsilon-near-zero hyperbolic metamaterials}

\abbreviations{IR,NMR,UV}

\begin{document}

\begin{tocentry}

Some journals require a graphical entry for the Table of Contents.
This should be laid out ``print ready'' so that the sizing of the
text is correct.

Inside the \texttt{tocentry} environment, the font used is Helvetica
8\,pt, as required by \emph{Journal of the American Chemical
Society}.

The surrounding frame is 9\,cm by 3.5\,cm, which is the maximum
permitted for  \emph{Journal of the American Chemical Society}
graphical table of content entries. The box will not resize if the
content is too big: instead it will overflow the edge of the box.

This box and the associated title will always be printed on a
separate page at the end of the document.

\end{tocentry}




\begin{abstract}

All-optical ultrafast switches enabled by artificial materials are considered at the forefront of the next generation of photonic communications and data processing. During the last two decades, the photonic applications, impact, and interest have tremendously increased in the framework of epsilon-near-zero (ENZ) photonics. Here, we experimentally propose a novel multilayered metamaterial utilizing Si-compatible  titanium nitride and indium-tin-oxide materials. The device exhibits two effective ENZ wavelengths in the visible and near-infrared spectrum, with switching times down to a few hundred femtoseconds at the corresponding ENZ regions. This novel approach will bring ENZ metamaterials towards new hybrid integrated CMOS photonic circuit components for ultrafast all-optical terahertz modulation.

\end{abstract}

\section{Introduction}
All-optical switches enable the ON/OFF conversion function by following the concept of light-controlled-by-light. They are highly significant due to their potential to overcome speed limitations set by electrical switches \cite{ono2020ultrafast,chai2017ultrafast,dhama2022all,caligiuri2020near}. Such an all-optical switching approach is the basis of the solid foundation for novel applications in optical communication networks, optical computing systems,  quantum information processing chips, and photonic central processing units \cite{vlasov2008high,leuthold2010nonlinear, shastri2021photonics}. This switching method mainly relies on third-order nonlinearities of optical materials, making these materials with large nonlinearities very crucial for ultrafast switching devices. In this regard, several different approaches based on photonic crystals \cite{scalora1994optical}, plasmonic metamaterials \cite{dani2009subpicosecond,bleckmann2014photochromic}, phase change materials \cite{choi2011nanopattern}, two-dimensional (2D) materials \cite{ono2020ultrafast}, hybrid perovskite materials\cite{stranks2013electron} and epsilon-near-zero (ENZ) materials\cite{kinsey2015epsilon,alam2016large} have been reported for the realization of ultrafast all-optical switches. ENZ materials, whose dielectric permittivity takes extremely low values \cite{engheta2013pursuing, vesseur2013experimental, wu2021epsilon} and eventually vanishes at a specific range of wavelengths are capable of modulating the refractive index at the femtosecond (fs) scale \cite{bohn2021all,vezzoli2018optical}. 

Recently, ENZ materials have especially emerged as a new paradigm for photonic platforms\cite{miscuglio2021approximate,amin2020sub,niu2018epsilon,galiffi2022photonics}. In particular, Si-compatible ENZ  materials that can be integrated with silicon platforms, such as photonics integrated circuits, are required to tremendously reduce the cost and manufacturing complexities allowing a more efficient performance and new functionalities in hybrid integrated photonics devices concepts\cite{minzioni2019roadmap,navarro2022ultrafast}.
In this respect, transparent conducting oxides (TCOs) which are complementary metal oxide semiconductor (CMOS) compatible, such as indium-tin-oxide (ITO) and aluminum-doped zinc oxide (AZO), turn out to be relevant homogeneous ENZ materials effective at the telecom window as ultrafast photonics components \cite{alam2016large,bohn2021all,kinsey2015epsilon}. However, these materials are limited by their intrinsic optical properties, and a tailorable ENZ is needed to enable ultrafast switches, particularly at the wavelengths of interest. 

Other than the homogeneous materials, subwavelength multilayer structures, often called hyperbolic metamaterials (HMMs) \cite{poddubny2013hyperbolic,guo2020hyperbolic,pianelli2022active}, are suitable candidates to design ENZ regime even towards optical wavelengths \cite{maas2013experimental,habib2020controlling}. Such layered structures have been envisioned as a potential new avenue for nonlinear optics because of their high nonlinear effective susceptibility\cite{boyd1994nonlinear}. The enhancement of nonlinear properties in the layered structures\cite{lepeshkin2004enhanced} and their nonlinearities have been theoretically analyzed in the framework of homogenized approximation \cite{ciattoni2010extreme}. There have been quite extensive works that mainly focus on 
the ultrafast nonlinear response in ENZ materials 
as well as effective ENZ multilayered structure like HMMs \cite{alam2016large,kinsey2015epsilon,rashed2020hot,kaipurath2016optically}.   
In particular, HMMs have proven to be excellent mediums for providing ultrafast optical switching in their effective ENZ \cite{kaipurath2016optically,rashed2020hot}. 
Yet, HMM-based on $Ag/SiO_{2}$ bilayers provides high nonlinear response with relatively fast ($\approx$ 1 ps) response times \cite{suresh2020enhanced}.
Recently, a third-order nonlinear optical response at the effective ENZ regime in an HMM composed by $Au/Al_{2}O_{3}$ bilayers was demonstrated 
with remarkably higher levels\cite{genchi2021tunable}.


Overall, to enable the integration of the ENZ materials with the photonic platforms, several drawbacks of ENZ materials need to be fulfilled for instance: (i) tailoring the effective ENZ at the wavelength of interest to achieve broadband or multiple wavelength ranges, (ii) the use of low-loss materials, and (iii) Si-compatible materials.\\ 
In this work, to address these criteria, we propose an HMM designed with two Si-compatible materials, i.e., titanium nitride (TiN) and ITO. The HMM with two bilayers possesses two-color ENZ, i.e., ENZ in visible and near-IR ranges (telecom windows spanning from O- to S-band). Using pump and probe transient absorption spectroscopy (TAS), we investigated the ultrafast responses of the HMM at its effective ENZ wavelengths, unveiling all-optical switching response times down to a few hundred femtoseconds (fs).
In particular, the HMM can simultaneously support multiple ultrafast switches at distinct ENZ wavelengths capable of all-optically modulating the response. Such remarkable novel all-optical modulation renders such metadevice an optimal candidate that unlocks new functionalities and possibilities in photonic integrated switches.

\section{Results and discussion}
As the first step, TiN and ITO layers were separately deposited by magnetron sputtering and electron beam evaporation on glass substrates, respectively. The optical constants (refractive index (n) and extinction coefficient (k)) of these films were retrieved by performing spectroscopic ellipsometry, as shown in Section 1 of the supporting information (SI).
These results have been further utilized in Ansys Lumerical software to perform numerical simulations in order to design the desired HMM.
In this context, two bilayers of TiN and ITO stacked films theoretically provide an HMM structure that enables two ENZ resonances. Thus, the final HMM structure has been realized by alternatively depositing the TiN (11 nm) and ITO (32 nm) up to two bilayers to obtain ENZ at two different wavelength regions, i.e., visible and near-infrared.

\begin{figure}[H] 
\centering
\includegraphics[width=0.9\textwidth]{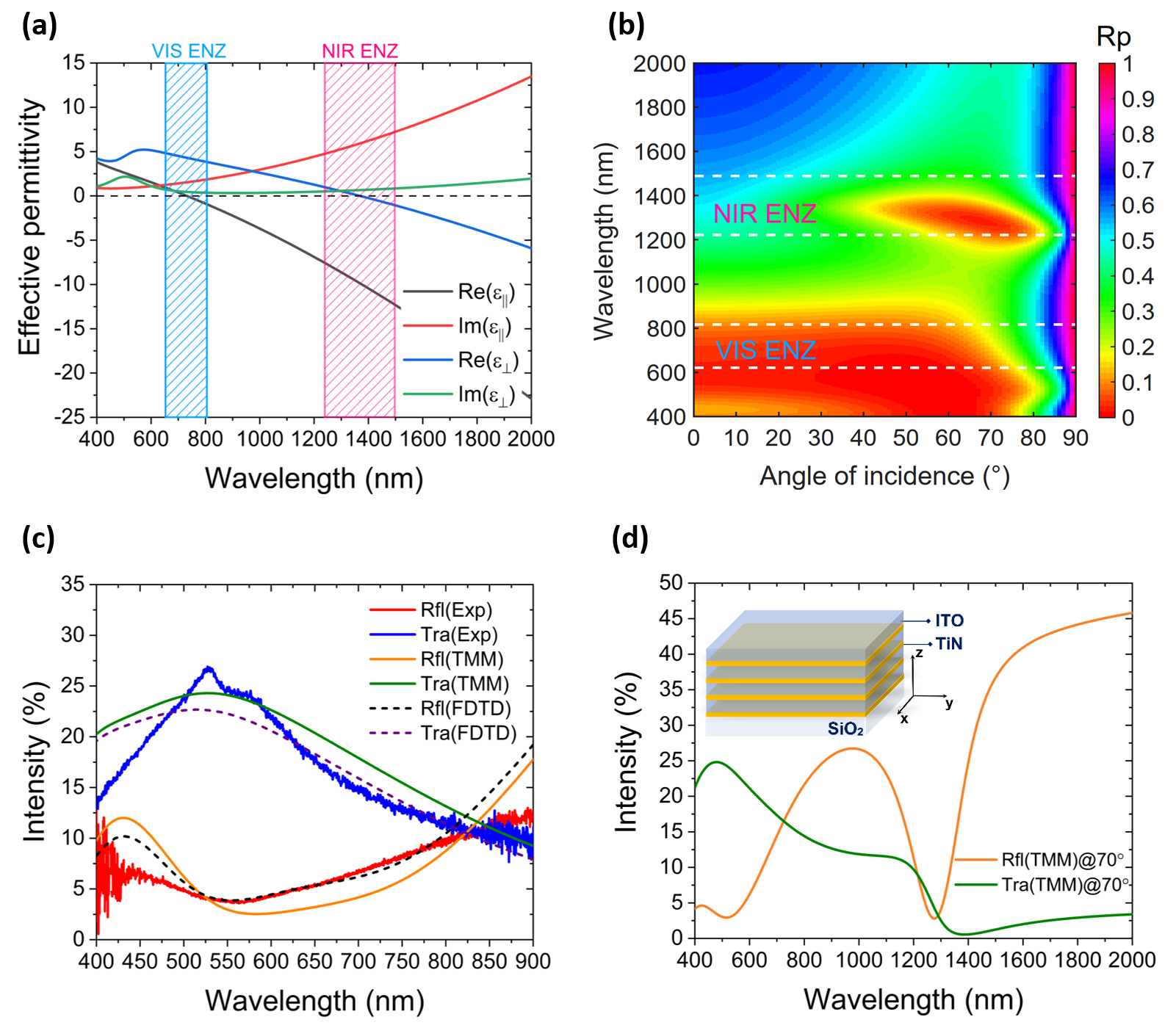}
\caption{(a) Effective real and imaginary parts of permittivities of TiN/ITO based HMM, (b) Calculated p-polarized
reflection spectrum via the TMM method showing ENZ region in visible (VIS) and near-IR (NIR) as a function of the angle of incident light ($\theta$).
(c) Experimental and numerically simulated reflection and transmission curves of HMM by using  FDTD and TMM methods. (d) Reflection and transmission curves at the excitation of $70^{\circ}$ angle of the incident light, while the inset depicts the sketch of the engineered HMM.} 
\label{fig:1}
\end{figure}

Figure \ref{fig:1} (a) shows the effective real and imaginary parts of the permittivities responses of the HMM in the visible and near-IR range. The wavelength region at which the parallel permittivity $\epsilon_{\parallel}$ crosses zero falls in the range of 649-810 nm and is referred to as the visible epsilon-near-zero (VIS ENZ) region. Additionally, the real part of the perpendicular permittivity $\epsilon_{\perp}$ vanishes in the range from 1238-1500 nm and is named the near-infrared ENZ (NIR ENZ) region. It is worth noting that the imaginary part of the permittivity in both ENZ remains low, particularly for the perpendicular imaginary part Im ${(\epsilon_{\perp})}$ in both ENZ regions. The appearance of these two ENZ regions is clearly shown in the  calculated reflection spectrum (see the dashed regions of Figure \ref{fig:1} (b)). One can observe that the VIS ENZ region is present at each incident angle till approx $70^{\circ}$ while the NIR ENZ region begins to appear when the HMM is illuminated with an impinging angle above $40^{\circ}$ degree, achieving  the maximum at an angle of $70^{\circ}$ degree. Furthermore, Figure \ref{fig:1} (c) shows good agreement of the HMM transmission and reflection curves between experimental results and simulations obtained via the Finite-difference time-domain (FDTD) and Transfer-matrix method (TMM).
While Figure \ref{fig:1} (d) presents the typical ENZ regime response with $70^{\circ}$ incident angle accordingly with the reflection distribution in Figure \ref{fig:1} (b). This response is also typically observed in ENZ materials films \cite{fruhling2022understanding}.

\subsection{Ultrafast switching response}

\begin{figure}[H] 
\centering
\includegraphics[width=0.98\textwidth]{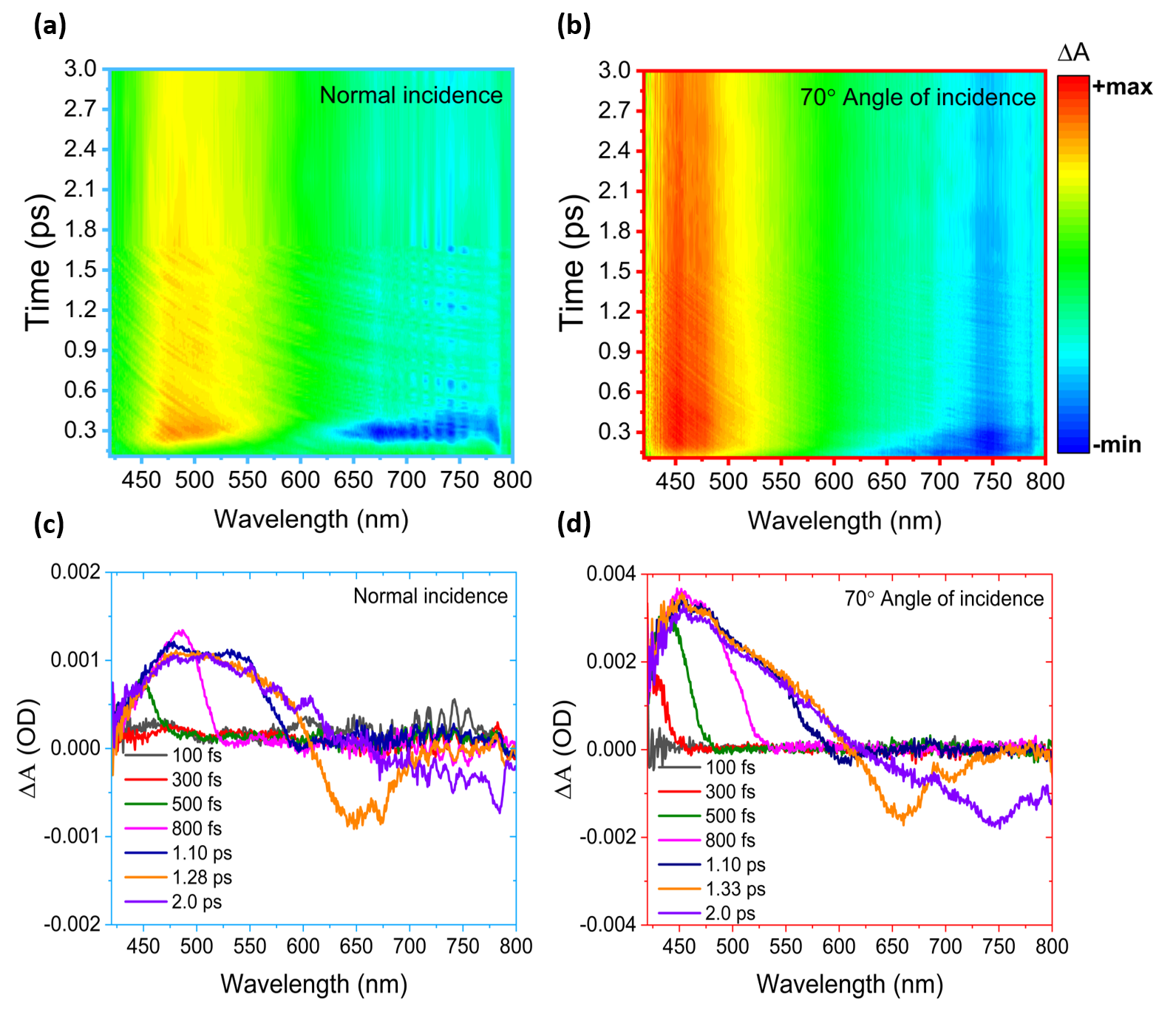}
\caption{2D color maps of transient absorption  as the function of wavelength and pump-probe delay time, when excited by 700 nm pump wavelength with a fluence of $1.40$ $mJ\cdot cm^{-2}$ (a) at the normal incidence and (b) at $70^{\circ}$ angle of the incident pump pulses. 
Transient absorption($\Delta$A (OD)) spectra curves at specific time delays(c) at normal incidence and (d) at $70^{\circ}$ angle of the incidence, extracted from (a) and (b), respectively.} 
\label{fig:2}
\end{figure}

In order to investigate the switching mechanism in VIS ENZ and NIR ENZ regimes, HMM was excited by intraband as well as interband pump wavelengths and probed in the visible and near-infrared regions utilizing transient absorption spectroscopy (TAS) in transmission mode. A pump beam with 100 fs pulse duration excites the system, and probe pulses are delayed with respect to pump pulses to ascertain the spatial and temporal dynamics of the two-color HMM in the wavelength region of interest. 
In particular, we emphasized transient modulation at intraband excitation in the corresponding ENZ regions as it induces highly non-thermal electrons, i.e., hot electrons (HEs) at lower energy consumption compared to interband excitation\cite{alam2016large,kinsey2019near} and enables relaxation dynamics of HEs at ultrafast sub-pico second scale. Furthermore, transient modulation experiments have also  been conducted  to investigate switching responses within one ENZ region upon the excitation of intraband excitation of other ENZ regions and vice versa.
To understand the change in switching time response of the VIS ENZ region due to the incident angle of the pump beam, we first excited the HMM at normal incidence and $70^{\circ}$ incidence angle by 700 nm (1.7 eV) pump wavelength with a fluence of $1.40$ $mJ\cdot cm^{-2}$ and probed in visible (420 - 800 nm) region. Figure \ref{fig:2} (a) and (b) show the 2D color maps of transient absorption (TA) as the function of wavelength and time at the normal incidence and $70^{\circ}$ angle of the incident pump pulses, respectively. Similarly, the induced  absorption modulation spectra are reported as a function of wavelength at different time delays, as shown in Figure \ref{fig:2} (c) and (d). As shown in Figure \ref{fig:2}, oblique ($70^{\circ}$ angle) incidence of pump pulses enables stronger transient response in comparison to normal incidence, and the system undergoes from excessive absorption's maxima (positive $\Delta$A) to negative transient absorption (bleach) region within sub-picosecond scale. Such an ultrafast transition from positive transient absorption to bleach region indicates the presence of optical switching at a few hundred femtoseconds time scale.

Three processes can mainly describe the physical phenomena behind such all-optical intraband ultrafast switching in HMM.
(i) In the absence of pump pulses, electrons of both building blocks are in the conduction band in their equilibrium state defined by Fermi distribution. (ii) After the pump pulse impinges the HMM, the electrons in the conduction band are energized, which quickly equilibrate via electron-electron (e-e) and electron-phonon (e-p) scattering, leading to the creation of a highly non-thermal distribution at a well-defined elevated temperature \cite{diroll2020broadband,PhysRevB.51.11433}.
(iii) Hot electrons dissipate energy through electron–phonon (e-p) interactions, increasing the lattice temperature until the system returns to its initial equilibrium state. This process is associated with the switching off-recovery time.
In general, the ultrafast relaxation time of HEs occurs at a rate mainly governed  by the electron-phonon (e-p) coupling over a timescale of femtoseconds to picoseconds and induced charge dynamics in the systems \cite{alam2016large,alam2018large,dhama2023unveiling}.

\begin{figure}[H] 
\centering
\includegraphics[width=0.98\textwidth]{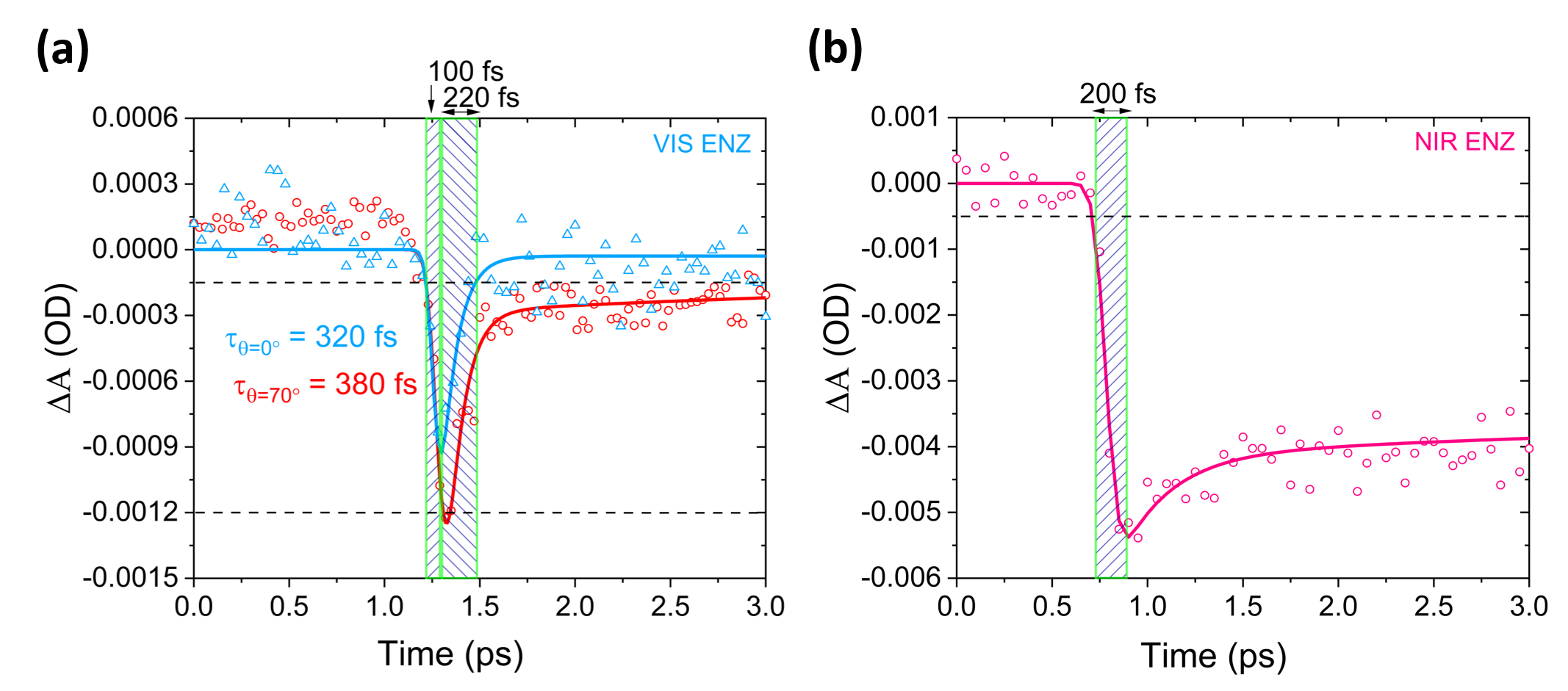}
\caption{Switching time responses of TiN/ITO based  HMM with excitation of 700 nm with a fluence of $1.40$ $mJ\cdot cm^{-2}$ (a) probed at 650 nm (VIS ENZ) for normal incidence and $70^{\circ}$ angle of incidence and (b) at 1240 nm (NIR ENZ) with $70^{\circ}$ angle of incidence.}  
\label{fig:3}
\end{figure}

We further studied the temporal dynamics of transient absorption modulation  of HMM in the VIS ENZ region at normal incidence and $70^{\circ}$ angle of incidence of pump and probe lights when excited by VIS ENZ wavelength (700 nm  pump pulses) under a similar pump fluence ($1.40$ $mJ\cdot cm^{-2}$). In this context, Figure \ref{fig:3} (a) visualizes the kinetic curves as the function of pump-probe delay time extracted from Figure \ref{fig:2} (a) and (b) at normal and oblique incidence to compare the switching time responses in both cases. Despite the stronger transient response under oblique incidence, HMM exhibits a faster switching time (blue curve) at normal incidence ($\tau_{\theta=0^{\circ}}$ = 320 fs) in comparison to the switching time (red curve) at $70^{\circ}$ angle of incidence ($\tau_{\theta=70^{\circ}}$ = 380 fs) for the transition wavelength of VIS ENZ region (650 nm), as shown in Figure \ref{fig:3} (a).



\begin{figure}[H] 
\centering
\includegraphics[width=0.98\textwidth]{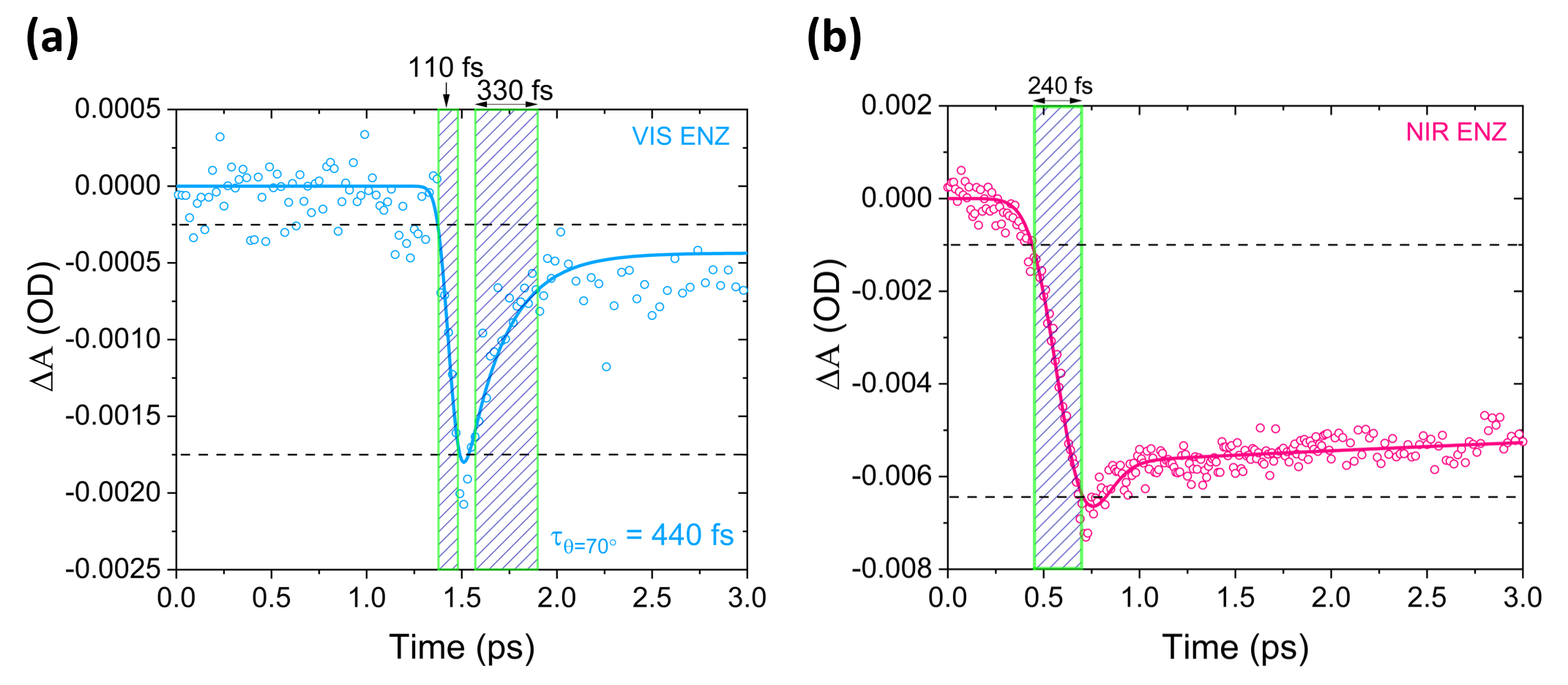}
\caption{Switching time responses of HMM with excitation of 400 nm pump pulses with a fluence of $1.40$ $mJ\cdot cm^{-2}$, at $70^{\circ}$ angle of incidence in (a) VIS ENZ at 650 nm probe wavelength and (b) NIR ENZ region at 1240 nm probe wavelength.} 
\label{fig:5}
\end{figure}

Furthermore, the temporal dynamics of transient response in the NIR ENZ region of HMM were investigated when excited at intraband excitation of the VIS ENZ region under similar experimental conditions. 
Figure \ref{fig:3} (b) reports a resolution-limited rise time of 200 fs, followed by a long-living plateau indicating longer relaxation dynamics of HEs. Fast time (200 fs) at sub-ps scale in such a dynamic behaviour is attributed to those HEs which relax through internal non-radiative relaxation processes, transforming from non-thermal to thermal energy distribution via electron–electron (e-e) scattering. Whereas the long-living plateau represents the slow dynamics due to the heat released from the lattice to the surrounding environment via phonon–phonon (p-p) scattering within a time scale from hundreds of picoseconds to nanoseconds\cite{schirato2023ultrafast}.
Indeed, if we expand the probe temporal delay time window to 200 ps,  
the heat of the lattice is mainly released via the phonon-phonon (p-p) scattering process with a relaxation time of $\approx$ 78 ps, as shown in Figure S\ref{fig:5} of Section 2.2 in the SI.


   \begin{figure}[H] 
\centering
\includegraphics[width=0.98\textwidth]{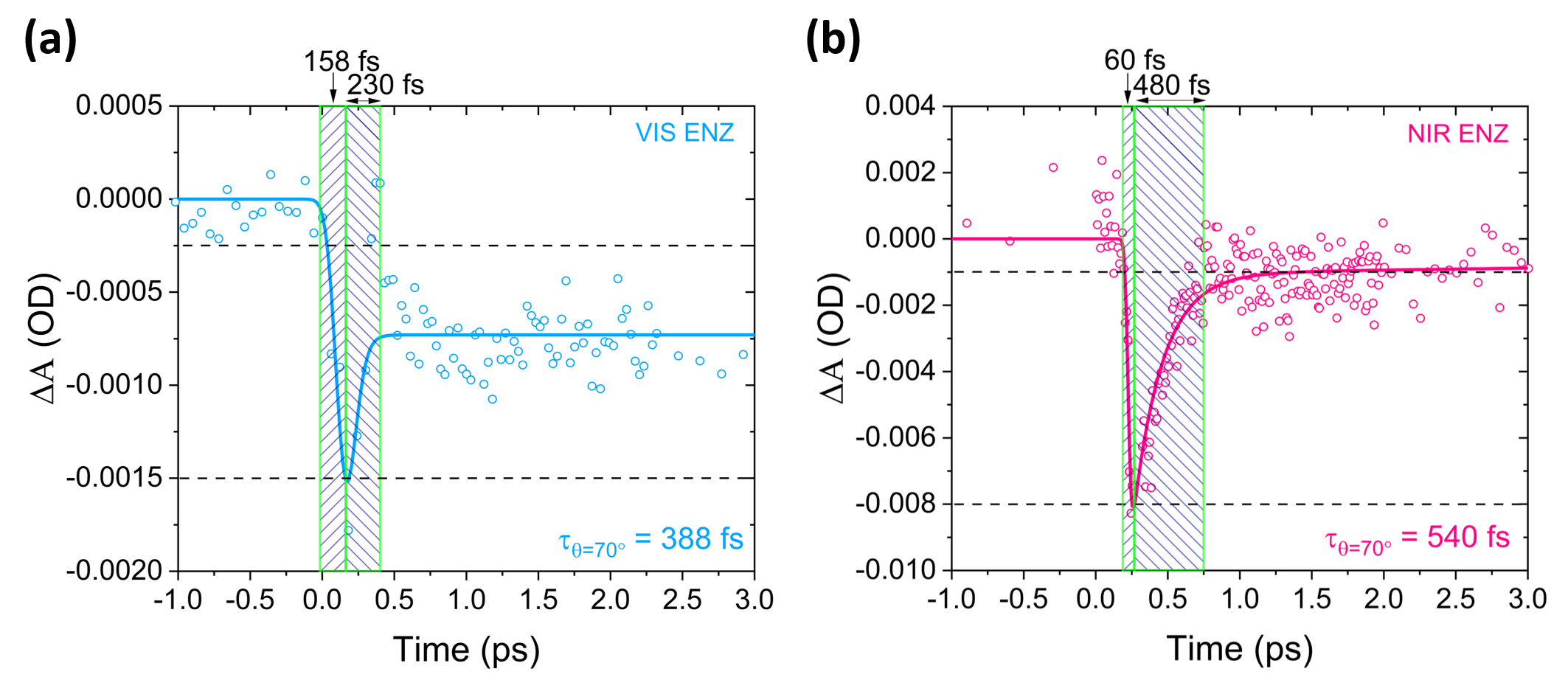}
\caption{Switching time responses of HMM by intraband excitation of NIR ENZ region (1250 nm) with the pump fluence of $1.16$ $mJ\cdot cm^{-2}$, at $70^{\circ}$ angle of incidence (a) in VIS ENZ at 783 nm probe wavelength and (b) NIR ENZ region at 1325 nm probe wavelength.} 
\label{fig:4}
\end{figure}


In addition, we investigated the switching performance of HMM in both ENZ regions when pumped at the intraband excitation wavelength of NIR ENZ region at $70^{\circ}$ angle of incidence. In this regard, Figure \ref{fig:4} (a) shows the temporal dynamics of transient modulation for 783 nm probe wavelength at the excitation of 1250 nm NIR pump pulses with the fluence of 1.16 $ mJ\cdot cm^{-2}$. 

The metadevice undergoes an ultrafast bleach response time of 158 fs, followed by a recovery process that takes 
388 fs to return to its partially recovered state as shown in Figure \ref{fig:4} (a). After this ultrafast temporal dynamics, a long-living plateau feature appears to remain constant for several hundred picoseconds. This behaviour is known as a "fast" and "slow" contribution of the relaxation time typically featured in TiN films in contrast to the gold switching dynamics \cite{george2019nonlinearities}. In particular, the observed slower temporal component suggests that the relaxation dynamics of the system are influenced significantly by the lattice temperature. This implies that the temperature of the lattice is an important factor in determining the temporal evolution of the system, especially when nitrides are involved\cite{george2019nonlinearities,diroll2020broadband}. 



Under the same pump fluence and similar intraband excitation (1250 nm NIR pump pulses), Figure \ref{fig:4}b demonstrates a remarkable optical switching performance of the metadevice in the NIR ENZ region when interrogated with 1325 nm probe wavelength. The bleach response maxima occur in $\approx$ 60 fs at an ultrafast time scale, followed by a recovery time of 480 fs within an overall temporal switching window of 540 fs. It is noted that our metadevice provides a faster switching response in its effective permittivity than that of homogeneous ENZ thin films and ENZ metasurface counterparts \cite{alam2016large,kinsey2015epsilon,diroll2020broadband,alam2018large}. In Figure S3 and S4, section 2.1 of the SI, we report the slower response time for a single layer of TiN (11 nm) and ITO (32 nm) films, respectively.

  



Finally, Figure \ref{fig:5} reports the transient switching response of HMM at the interband excitation under the illumination of 400 nm pump pulses.
In this context, Figure \ref{fig:5} (a) presents the reasonable switching response with a recovery time down to 440 fs at 650 nm probe wavelength in  the VIS ENZ region.

While the device does not exhibit any switching response at an ultrafast time scale (within 3 ps temporal window) in the NIR ENZ region, as demonstrated in Figure \ref{fig:5} (b). In this scenario, as the system is interrogated using high-energy pump photons, the lattice temperature rises very fast due to the high scattering rate of the higher energy electrons followed by a slow long-lasting thermal relaxation of the HEs\cite{schirato2023ultrafast, norris2003femtosecond,george2019nonlinearities}. At a longer pump-probe temporal window scale, the metadevice possesses a longer recovery time of $\approx$ 110 ps, as shown in Figure S6, section 2.2 in the SI. 

We summarized the metadevice switching performance in VIS and NIR ENZ regions for different excitation wavelengths, as shown below in Table 1. 








\begin{table*}
  \centering
  \begin{tabular}{|P{2cm}|P{2cm}|P{2cm}|P{2.5cm}|P{2.5cm}|P{2.5cm}|}
    \hline
    Pump $\lambda$ (nm) & Probe region & Switching time ($\tau$) @ ENZ & Angle of Incident ($\theta$) & Intensity recovery (\%) & Fluence ($mJ\cdot cm^{-2}$) \\
    \hline    
     400   & VIS  & 440 fs  & $70^{\circ}$      & 80  & 1.40 \\
    400   & NIR  & 110 ps & $70^{\circ}$      & 72  & 1.40 \\
    700   & VIS  & 320 fs & $0^{\circ}$       & $\approx$ 100 & 1.40 \\
    700   & NIR  & 78 ps  & $70^{\circ}$      & 76  & 1.40 \\
     1250  & VIS  & 388 fs & $70^{\circ}$      & 50  & 1.16 \\
    1250  & NIR  & 540 fs & $70^{\circ}$      & 90  & 1.16 \\
    \hline
  \end{tabular}
  \caption{Two-color all-optical switching performance of HMM with TiN/ITO layers}
\end{table*}

\subsection{Conclusions}

In conclusion, we reported ultrafast switching response in two effective ENZ regions of an HMM based on low-loss and CMOS-compatible building blocks of TiN/ITO bilayers. The results show that the hot electrons dynamics effects are faster in the effective near zero permittivity region of the HMM rather than its thin homogeneous layer counterpart, i.e., TiN or ITO film. This opens up a new approach towards realizing tunable and reversible ultrafast all-optical switching at the fs scale.
Taking advantage of low losses and excellent optical properties of TiN and ITO at studied wavelengths, we observed fast switching time in VIS as well as NIR ENZ wavelength regions. This metadevice can function as a double-switching device featuring an all-optical modulation response within the 1.8 - 3.1 THz range. Our results envision the possibility of particular engineering metasurfaces that can potentially be used as fast modulators operating at optical wavelengths. Such metamaterials have the potential to launch a new paradigm towards hybrid configuration for ultrafast modulators and photonic-integrated CMOS chips. Such proposed Si-compatible metamaterials can also provide an optimal trade-off between modulation speed and efficient performance in silicon-based photonic systems.


\subsection{Sample preparation}
Titanium nitride (TiN) thin films were deposited by magnetron sputtering technique using Oxford Instruments Plasmalab System 400. Pure Titanium (Ti) was sputtered using a mixture of argon and nitrogen. The addition of nitrogen made the process reactive. The film composition, controlled by the ratio of the argon to nitrogen gas flow, was chosen to be stoichiometric since the plasmonic properties, e.g., the plasmonic figure of merit, are composition-dependent. The process parameters – applied power, pressure, and gas flows were chosen to provide films with the highest value of the figure of merit. All layers were deposited at a temperature of $21^{\circ}C$ with chamber base pressure below $5\times10^{-7}$ Torr. The deposition time was 11 seconds to get 11 nm thick films. Indium-tin-oxide (ITO) was prepared by electron beam evaporation in a custom-made vacuum chamber at the base pressure of $1\times10^{-6}$ mbar. Physical vapor deposition processes obtained ITO layers. For the evaporation of these layers, a vacuum chamber equipped with a telemark 264 electron gun as an evaporation material source, quartz weight for controlling layer evaporation rate and thickness, a heating system for regulating substrates temperature, the system for dosing of oxygen pressure application during evaporation process were used. The ITO deposition process was carried out for $200^{\circ}C$ substrate temperature, under oxygen pressure $7\times10^{-5}$ mbar. ITO material for evaporation consisted of $10\%$ $SnO_{2}$ to $90\%$ $In_{2}O_{3}$ by weight.
\subsection{Optical measurements}
The reflectance and transmittance measurements were carried out using a broadband optical source (Energetiq EQ-99XFC LDLS) to excite the samples. The optical spot is focused on the samples using a Zeiss "Epiplan-Neofluar" 20X objective (NA=0.4) for both reflectance and transmittance measurements. 
The response of the samples was coupled to an optical fiber connected to Ocean Optics Flame UV-VIS spectrometer for the spectral response.

\subsection{Transient absorption spectroscopy } 

Ultrafast time-resolved pump-probe spectroscopic measurements were performed using an amplified Ti: sapphire laser system equipped with an optical parametric amplifier (OPA). This system produced 100 fs pulses at 1.00 kHz with a center wavelength of 800 nm. Most of the output power (90$\%$) was sent to OPA to generate tunable pump pulses in the UV-Visible to near-infrared spectral regions to excite the samples at the desired wavelength. The remaining 10$\%$ of output power travels through a delay line, enabling a controlled time difference between pump and probe pulses. It converts into a broadband probe beam to interrogate the sample in transmission mode. At the same time, the chopper-modulated pump pulse is spectrally as well as temporally overlapped with the probe beam on the sample. At the same time, the detector is triggered to detect every probe pulse and calculate the absorption spectrum. Repetition rates of the pump and probe beams are 500 Hz and 1 kHz, respectively. Therefore, the effect of the pump beam will be observed only in one of the two consecutive probe beams.


\begin{acknowledgement}

We acknowledge the financial support of the European Research Council (Starting Grant project aQUARiUM; Agreement No. 802986), Academy of Finland Flagship Programme (PREIN), (320165). All authors thank Dr. Urszula Chodorow for her support in ellipsometric measurements. J. J. thanks to the POB FOTECH-3 project entitled "Plasmons and polaritons on nanostructured surfaces of IVb metal nitrides" granted by the Warsaw University of Technology within "The Excellence Initiative – Research University" program.

\end{acknowledgement}




\bibliography{acs-achemso}

\end{document}